\documentclass[prl,twocolumn,10pt,showpacs,aps]{revtex4-1}
\usepackage{graphicx,amsfonts,times,bm,amsmath,amsfonts,verbatim,soul}
\newcommand{\ket}[1]{|#1\rangle}

\begin{document}

\title{Hidden Symmetry Decoupling of Majorana Fermions}

\author{Eugene Dumitrescu$^1$, Tudor D. Stanescu$^2$, Sumanta Tewari$^1$}
\affiliation{$^1$Department of Physics and Astronomy, Clemson University, Clemson, SC 29634, USA\\
$^2$Department of Physics, West Virginia University, Morgantown, WV 26506}
\begin{abstract}
Multiple zero-energy Majorana fermions (MFs) with spatially overlapping wave functions can survive only if their splitting is prevented by an underlying symmetry. Here we show that, in quasi-one-dimensional (Q1D) time reversal invariant topological superconductors (class DIII), a realistic model for superconducting
Lithium molybdenum purple bronze (Li$_{0.9}$Mo$_6$O$_{17}$) and certain families of organic superconductors, multiple Majorana-Kramers pairs with strongly overlapping wave functions persist at zero energy even in the absence of an easily identifiable symmetry. We find that similar results hold in the case of Q1D semiconductor-superconductor heterostructures (class D) with $t_{\perp} \ll t$, where $t_{\perp}$ and  $t$ are the transverse and longitudinal hoppings, respectively. Our results, explained in terms of special properties of the Hamiltonian and wave functions, underscore the importance of hidden accidental symmetries in topological superconductors.
\end{abstract}

\pacs{03.65.Vf, 71.10.Pm, 03.67.Lx}
\maketitle

\textit{Introduction:}
Topological superconductors are characterized by a bulk superconducting gap and topologically protected gapless edge states \cite{Hasan_2010}. Due to the presence of intrinsic superconducting particle-hole symmetry (PHS) the gapless zero modes constitute Majorana fermions (MFs), characterized by the second quantized operator relation $\gamma^{\dagger}=\gamma$. In the context of condensed matter, aside from being fascinating non-elementary particles, MFs obey Ising type non-Abelian braiding statistics which is useful in implementing a fault-tolerant topological quantum computer \cite{Nayak_2008}.
These emergent excitations are said to be topologically protected, in the sense that their existence and properties are insensitive to many perturbations so long as the system remains gapped.
While MFs have not yet been conclusively found in nature, they have been theoretically shown to exist in low dimensional spinless $p$-wave superconducting systems \cite{Read_Green_2000,Kitaev_2001} as well as other systems which are similar to them \cite{Fu_2008,Zhang_2008,Sato_2009,Sau,Long-PRB,Roman,Oreg}.
In particular the semiconductor heterostructure scheme has motivated tremendous experimental efforts with a number of recent works claiming to have observed experimental signatures consistent with MFs \cite{Mourik_2012,Deng_2012,Das_2012,Rokhinson_2012,Churchill_2013,Finck_2013}, for a review see Ref.~[\onlinecite{Stanescu_2013}].

 Recent work \cite{Schnyder_2008,Kitaev_2009,Ryu_2010} has established that the quadratic Hamiltonians for gapped topological insulators and superconductors can be classified into ten topological symmetry classes each of which is characterized by a topological invariant.
The symmetry classification is important as it provides an understanding of the effects of various perturbations on the stability of the protected surface modes such as MFs. For example, recent work \cite{Law_12,Deng_12,Keselman_13,Zhang_13,Deng_13,Law_13,Flensberg_13,Nakosai_13,Dumitrescu_13,Dumitrescu2_13} has proposed time-reversal (TR) invariant topological superconductivity (class DIII) with a $\mathbb{Z}_2$ invariant in a number of systems with intrinsic or proximity induced superconductivity in  heterostructures. Spin triplet, equal spin pairing (ESP), $p$-wave superconductivity, which realizes such a TR-invariant topological superconductor \cite{Dumitrescu_13}, is thought to be present in the quasi-one-dimensional (Q1D) transition metal oxide Lithium molybdenum purple bronze Li$_{0.9}$ Mo$_6$ O$_{17}$ (LiMO) and some organic superconductors \cite{Lebed_00,Mercure_12,Lebed_13,Lee_2001,Lee_2003,Shinagawa_2007}. These systems posses a distinctly anisotropic electrical conductivity, i.e. the hopping integrals along the crystallographic directions vary as, $t_x \gg t_y \gg t_z$, making them Q1D conductors. Because of its electronic anisotropy, LiMO may be modeled as an array of parallel one dimensional systems coupled by weak transverse hopping. In principle, such weakly coupled array of parallel one dimensional topological superfluids can also be realized in cold fermion systems \cite{Qu_14}.  As discussed below these systems provide a natural platform to study interaction effects between MFs.

In Q1D multi-chain systems multiple Majorana fermions with spatially overlapping wave functions can remain at zero energy only if their splitting is forbidden by an underlying symmetry. In this work we show that, in Q1D TR-invariant topological superconductors, multiple Majorana-Kramers pairs (MKPs) with strongly overlapping wave functions persist at zero energy even in the absence of an identifiable physical symmetry. We find similar results also
 for Q1D semiconductor-superconductor heterostructures with spin-orbit (SO) coupling and Zeeman field (class D with $\mathbb{Z}_2$ invariant) with $t_{\perp} \ll t$, pointing to the existence of a hidden symmetry decoupling of the MFs.
To demonstrate this result we start with a strictly 1D (single chain) TR-invariant Kitaev model superconductor (Eq.~(\ref{eq:H1D})), modeling the ESP spin-triplet $p$-wave state proposed to be realized in LiMO.
We first note that, in addition to TR-invariance, the model has a chiral as well as a mirror symmetry both of which allow an integer ($\mathbb{Z}$) invariant. In the physically realistic Q1D generalization of this model
(with $t_{\perp} \ll t_x$) the $\mathbb{Z}$ invariant takes arbitrary integer values,  allowing multiple MKPs localized at the same end despite wave-function overlap.
We show that, even in the absence of such symmetries, multiple MFs can still be protected by symmetries such as
spatial reflection. In realistic materials, however, reflection symmetry is expected to be broken by disorder. Remarkably, we find that disorder induced breakdown of reflection symmetry fails to lift the degeneracy of the zero energy modes even with strong wave function overlap. We find very similar results also for Q1D systems in class D. These results, which we explain in terms of special properties of the Hamiltonians and wave functions, underscore the importance of hidden symmetry decoupling of MFs in topological superconductors.

\textit{Hamiltonian and equivalent description by chiral and mirror symmetries:}
 We model a one dimensional spin-triplet topological superconductor by a lattice Hamiltonian which includes nearest neighbor hopping, on-site chemical potential and a general $p$-wave superconducting order parameter which reads,
\begin{eqnarray}
H^{\text{1D}}&=&\sum_{i,\sigma,\sigma'}[-t c_{i+1\sigma}^{\dagger}c_{i,\sigma}^{  }-\mu c_{i\sigma}^{\dagger}c_{i\sigma}^{ } \nonumber \\
&+& \Delta_{\sigma \sigma'}(c_{i+1\sigma}^{\dagger}c_{i\sigma'}^{\dagger} +H.c.)].
\label{eq:H1D}
\end{eqnarray}
Here $t=t_x$ is the hopping integral between nearest neighbor sites, $i \in [1,N_x]$ is the lattice index and $\sigma=\uparrow,\downarrow$ represents the spin index. Next, we Fourier transform Eq. (\ref{eq:H1D}) to study the 1D bulk topological properties.
In momentum space, the superconducting gap function which describes correlations between electrons is written $\Delta_{\alpha\beta}(\textbf{k})=\left\langle c_{\alpha}(\textbf{k})c_{\beta}(-\textbf{k}) \right\rangle$, where $c_{\alpha}(\textbf{k})$ is the destruction operator of a single electron with spin $\alpha$ and momentum $\textbf{k}$.
The spin symmetry of Cooper pairing may be classified by the total spin as either singlet ($S=0$) or triplet ($S=1$). A general pair potential is expressed compactly in terms of a $d$-vector as $\Delta_{\alpha\beta}(\textbf{k})=\left[\Delta_s(\textbf{k})+\textbf{d}(\textbf{k})\cdot \bm{\sigma} \right](i\sigma_y)_{\alpha\beta}$ with a symmetric singlet component $\Delta_s(\textbf{k})=\Delta_s(-\textbf{k})$, and an antisymmetric triplet $d$-vector $\textbf{d}(\textbf{k})=-\textbf{d}(-\textbf{k})$.
In this work we will consider a pure triplet order parameter with $\Delta_s(\textbf{k})=0$, however the following analysis is general and applicable in the presence of a singlet term.
As an example, consider a Cooper pair in a state with zero spin projection along $\hat{z}$, that is $S_z = 0$. This is in fact a Cooper pair described by the familiar triplet state $\ket{\uparrow \downarrow}+\ket{\downarrow \uparrow}$ and corresponds to a pairing potential in Eq. \ref{eq:H1D} with $\Delta_{\uparrow,\downarrow}=\Delta_{\downarrow,\uparrow}$.

We now write the bulk Hamiltonian corresponding to Eq. (\ref{eq:H1D}) as a momentum space Bogoliubov-de Gennes (BdG) Hamiltonian, $H^{\text{1D}}=\sum_{k}\Psi_{k}^{\dagger} {\cal{H}}_{k} \Psi_{k}$. With the Nambu basis $\Psi_{k}=(c_{k\uparrow},c_{k\downarrow},c_{-k\downarrow}^{\dagger},-c_{-k\uparrow}^{\dagger})^{T}$ which absorbs the factor $i\sigma_y$ associated with the $d$-vector, the matrix ${\cal{H}}_{k}$ then takes the form
\begin{eqnarray}
\label{eq:H1DK}
{\cal{H}}^{1D}_{k}&=&(\epsilon(k)-\mu)\sigma_{0}\tau_z + \textbf{d}(k) \cdot \bm{\sigma} \tau_{x}.
\end{eqnarray}
Here $k=k_x$ is the 1D crystal-momentum, $\epsilon(k)=-2t(\cos(k)-1)$ is the single particle kinetic energy, $\textbf{d}(k)= \hat{d}|\textbf{d}(k)|= (d_x,d_y,d_z)\Delta\sin(k)$ is the $p$-wave order parameter and $\sigma_{i}$,$\tau_{i}$ indicate spin $1/2$ Pauli matrices in the spin and the particle-hole spaces respectively. The bulk spectrum consists of two doubly degenerate bands given by the dispersion relation $E_\pm=\pm \sqrt{(\epsilon_{k}-\mu)^2 +|\textbf{d}(k)|^2}$.

Superconducting Hamiltonians observe an intrinsic particle-hole symmetry (PHS) which emerges from the structure of the BdG equations. The BdG Hamiltonian in Eq.~(\ref{eq:H1DK}) satisfies
$\Xi {\cal{H}}_{k} \Xi^{-1}=-{\cal{H}}_{-k}$, where,
in this basis the anti-unitary PHS operator reads $\Xi=\sigma_y\tau_y{\cal{K}}$ where ${\cal{K}}$ is the anti-unitary complex conjugation operator. $\Xi$ anti-commutes with the real space representation of ${\cal{H}}^{1D}_{k}$ and obeys $\Xi^2=1$.
PHS which relates quasiparticle excitations at $\pm E$ through $\Gamma^{\dagger}_{E}=\Gamma_{-E}$ is fundamentally important for the formation of Majorana modes which are a special case satisfying $E=0$.
Additionally, ${\cal{H}}^{1D}_{k}$ obeys the time-reversal symmetry (TRS) relation
$
\Theta {\cal{H}}_{k} \Theta^{-1}={\cal{H}}_{-k}
$
with the TR operator $\Theta=\sigma_y\tau_0{\cal{K}}$.
The presence of PHS and TRS leads to a unitary chiral symmetry which is simply the product $\Pi=\Xi \cdot \Theta=\sigma_0 \tau_y$.
When $|\mu|<2t$ the system is in the topologically non-trivial phase, characterized by a DIII class $\mathbb{Z}_2$ invariant which takes a value $\nu=-1$. This invariant may be viewed as a Kramers polarization and reduces to Kitaev's Pfaffian invariant for one spin block in the presence of spin rotation symmetry \cite{Budich_13}. When $\nu=-1$ unpaired MFs at each end of the wire form topologically protected MKP's. This explains the robustness of the four zero energy modes in the presence of TR-invariant perturbations, for example spin-orbit coupling terms
such as ${\cal{H}}^{SO}=\alpha_R \sin(k) \sigma_y \tau_z$ added to Eq.~(\ref{eq:H1DK}).
\begin{figure}[t]
\includegraphics[width=4.5cm]{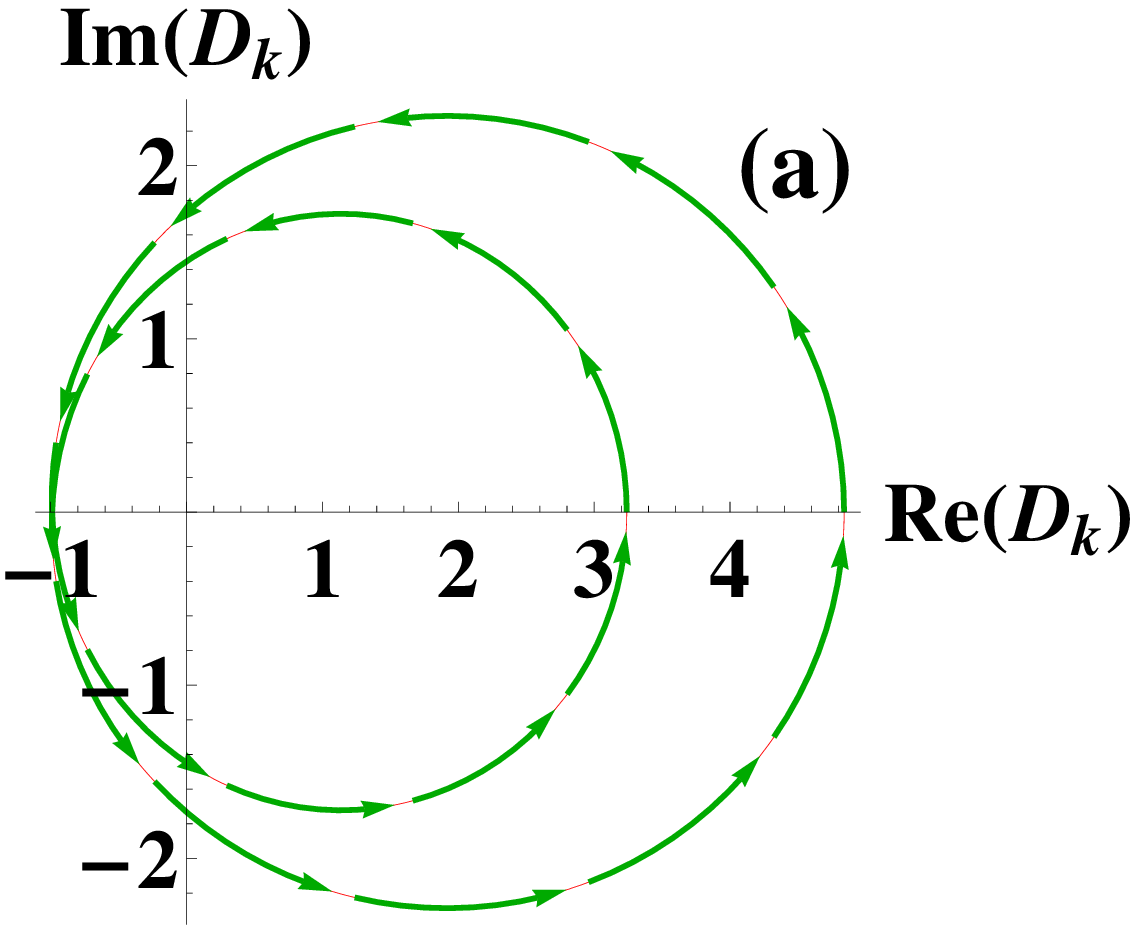} \includegraphics[width=4cm]{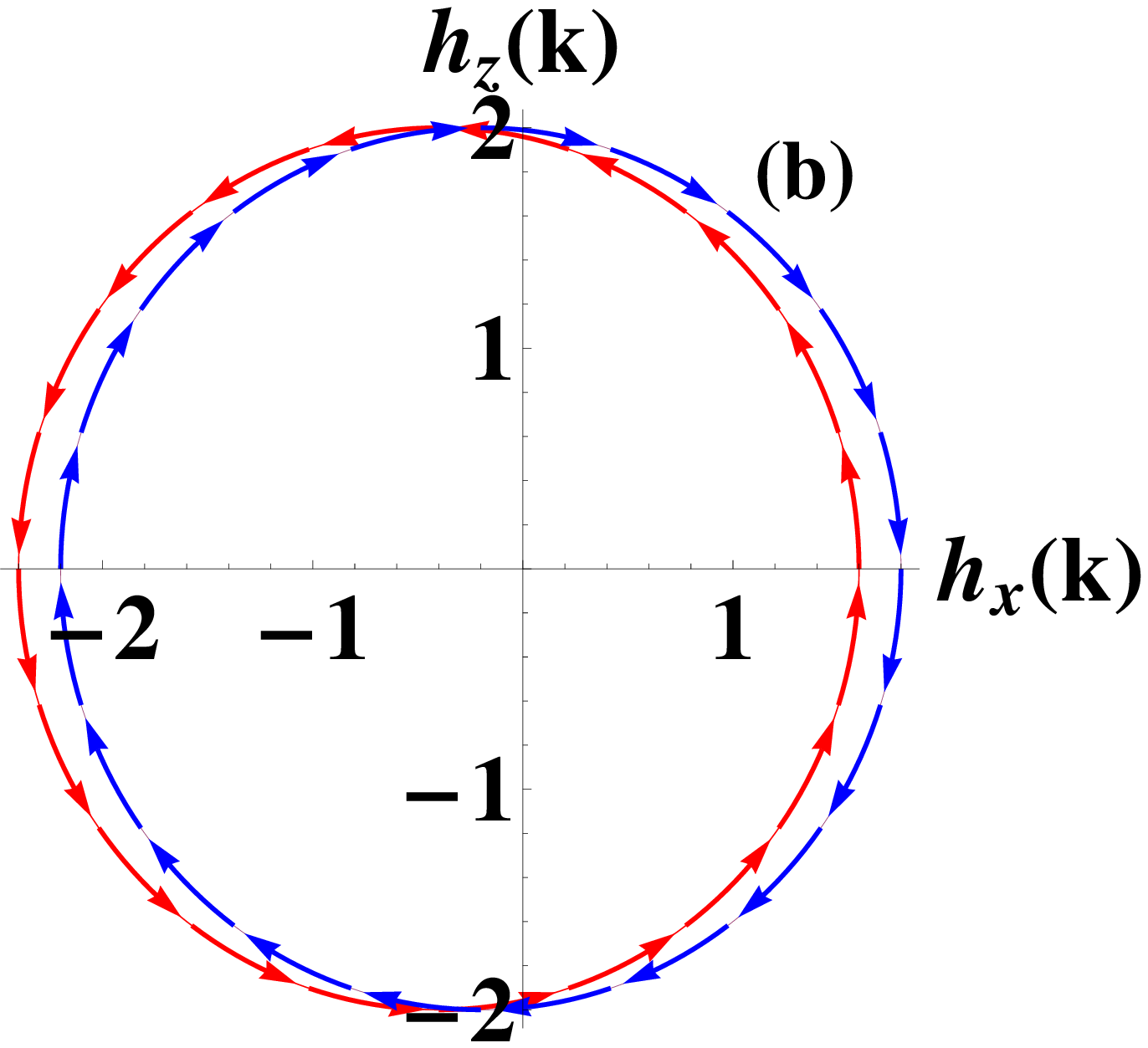}
\caption{\label{fig:BDI_Mirror} (Color online) (a) Chiral topological invariant $W=2$ indicating two topologically protected MFs at each end of a single chain described by Eq.~(\ref{eq:H1DK})
(b) Mirror topological invariant $\gamma_M=2$ (difference of the winding numbers in the two mirror subsectors) also indicating two MFs at each end of a single chain in Eq.~(\ref{eq:H1DK}).
}
\end{figure}

Additionally, the BdG Hamiltonian Eq.~(\ref{eq:H1DK}) belongs to the topological class BDI, due to a co-existing chiral symmetry given by ${\cal{S}}^{BDI}={\cal{O}}\cdot{\Xi}=(\hat{d}\cdot \bm{\sigma})\tau_{y}$, which is the product of a TR-like operator ${\cal O}=(\hat{d}\cdot\hat{y}+i(\hat{d}\times\hat{y})\cdot\bm{\sigma}){\cal K}$
with ${\cal{O}}^{2}=1$, and the particle-hole operator $\Xi$. In $d=1$ BDI Hamiltonians are classified by a bulk $\mathbb{Z}$ topological winding number invariant $W$.
To calculate the invariant we off-diagonalize the Hamiltonian from Eq.~(\ref{eq:H1DK}) in the basis which diagonalizes ${\cal{S}}^{BDI}$.
 Writing the determinant of the off-diagonal part in a complex polar form, $D_k=|Det(D_k)|e^{i\theta(k)}$, $W$ is given by \cite{Tewari_PRL_2012,Tewari_PRB_2012}
the number of times $\theta(k)$ winds about the origin as $k$ varies through the 1D Brillouin zone.
As can be seen from Fig. [\ref{fig:BDI_Mirror}], panel (a), the invariant takes the value $W=2$ in the topological phase of Eq.~(\ref{eq:H1DK})  while $W=0$ in the trivial phase.
This chiral symmetry explains the persistence of the zero modes to TR-breaking terms including stray Zeeman fields (${\cal{H}}^{Z}=\bm{V} \cdot \bm{\sigma} \tau_0$) perpendicular to the $d$-vector.
A generic SO coupling term aligned in an arbitrary direction in spin space is written ${\cal{H}}^{SO}=\alpha_R \sin(k) \bm{a} \cdot \bm{\sigma} \tau_z$ meaning that $\bm{a}\parallel\bm{d}_k$ preserves chiral symmetry while a SO term in the plane perpendicular to the $d$-vector does not respect chiral symmetry.

Recently, mirror symmetry has also been a proposed as a topological protection mechanism for MFs \cite{F_Zhang_13}.
The one-dimensional Hamiltonian Eq.~(\ref{eq:H1DK}) is invariant ($\left[{\cal{M}},{\cal{H}}_k\right]=0$) under the mirror symmetry operator ${\cal{M}}=i \hat{d}\cdot \bm{\sigma} \tau_0$.
Because ${\cal{M}}$ and ${\cal{H}}_k$ commute, the Hamiltonian may be expressed in a block diagonal form where each block corresponds to a mirror eigenspace subsector. Each block is written $\bm{h}^{\pm}\cdot \bm{\sigma}$ where the $\pm$ is the mirror eigenspace index. Explicitly choosing $\hat{d}$ along $\hat{x}$ we find $\bm{h}^{\pm}=(\pm \Delta \sin(k),0,\epsilon_k-\mu)$ such that the mirror winding number invariant in each subsector $C^{\pm}$ is defined in the $(\sigma_x-\sigma_z)$ plane. Each mirror winding curve encloses the origin once, but with opposite helicity, leading to a mirror invariant $\gamma_M=C^{+}-C^{-}=2$, which is illustrated in Fig. \ref{fig:BDI_Mirror}.

\textit{Multiple chains and Majorana multiplets:}
A realistic quasi-1D spin triplet superconductor such as LiMO (or quasi-1D TRI systems in cold fermions) may first be modeled as an array of 1D chains coupled by a weak hopping amplitude $t_y\ll t _x$. One may further consider a truly 3D system by stacking 2D arrays and coupling them through a third hopping integral $t_z\ll t_y \ll t_x$.
We consider a system which consists of $N_y$ parallel chains, indexed by $l \in [1,N_y]$, coupled only by transverse hopping $t_y$.
The quasi-1D Hamiltonian is a generalization of Eq.~(\ref{eq:H1DK}) given by
\begin{equation}
H^{Q1D}=\sum_{k l l'} \Psi_{k l}^{\dagger} ({\cal{H}}^{1\text{D}}_{k} \delta_{l,l'}+{\cal{H}}^{\perp}_{l,l'}) \Psi_{k l'},
\label{eq:HQ1D}
\end{equation}
where we have used the basis $\Psi_{k l}=(c_{k l \uparrow},c_{k l \downarrow},c_{-k l \downarrow}^{\dagger},-c_{-k l \uparrow}^{\dagger})^{T}$, and
${\cal{H}}^{\perp}_{l,l'} = - t_\perp \sigma_0 \tau_z  (\delta_{l,l'+1}+\delta_{l,l'-1})$.

We proceed by first examining a double chain setup with $l=1,2$ as an illustrative example. The generalization to multiple chains should be straightforward.
For a two-chain system the Hamiltonian is expressed as $2\times 2$ matrix where every entry is itself a $4\times 4$ matrix (see Eq. \ref{eq:H1DK}). This reads,
\begin{equation}
\label{eq:HQ1D}
H^{\text{Q1D}}_{}=\sum_{k}(\Psi^{\dagger}_{k,1},\Psi^{\dagger}_{k,2})
\begin{pmatrix}
  {\cal{H}}^{1D}_k & -t_\perp \sigma_0 \tau_z \\
  -t_\perp \sigma_0 \tau_z & {\cal{H}}^{1D}_k
 \end{pmatrix}
\begin{pmatrix}
  \Psi^{}_{k,1}  \\
  \Psi^{}_{k,2}
 \end{pmatrix}
\end{equation}

\begin{figure}[t!]
\includegraphics[width=6cm]{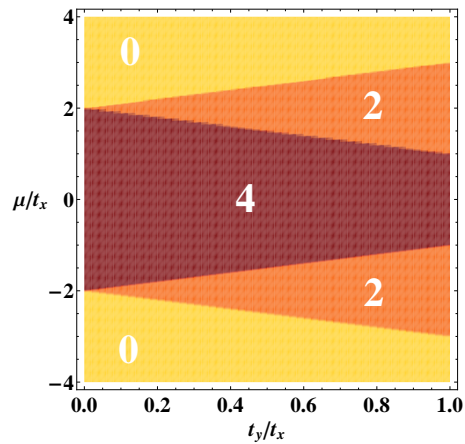}
\caption{\label{fig:BDI_Phase_Diagram} (Color online)
Phase diagram for a double chain set-up of TR-invariant Kitaev system (or generic class DIII superconductors) coupled by weak transverse hopping $t_y \ll t_x$. A large range of $\mu$ accommodates the topological phase indexed by the topological invariant $|W|=4$ which counts the number of localized Majorana modes at each end.}
\end{figure}

Introducing a new Pauli matrix ($\rho$) in the double chain Hilbert space allows us to write Eq. (\ref{eq:HQ1D}) compactly as ${\cal{H}}^{1D}_k \rho_0+t_\perp \sigma_0 \tau_z \rho_x$.
Using this, we may generalize the chiral operator to the double chain space as ${\cal{S}}^{BDI}=\sigma_{x}\tau_{y} \rho_0$.
We are now able to calculate a generalized multi-chain winding number $W$ counting the number of MFs at each edge which are now localized across both chains. Just as in the single chain case, the magnitude of the chiral invariant $|W|$ is equal to the number of topologically protected MFs present at each end. This is illustrated by the phase diagram presented in Fig. \ref{fig:BDI_Phase_Diagram} which shows that for small transverse hopping $2\times N_y=4$ Majorana modes are present.
In general, as long as one can define a chiral and/or mirror invariant, and the transverse hopping is small enough, the number of MFs at each end grows with the size of the sample ($|W|=2\times N_y$).

Let us now investigate the fate of Majorana multiplets in the event of broken chiral and mirror symmetries, which may occur due to intra-chain spin-orbit coupling perpendicular to the $d$-vector.
This modifies the Hamiltonian in Eq.~(\ref{eq:H1DK}) to,
$${\cal{H}}^{1D}_{k} \mapsto {\cal{H}}^{1D}_{k}+\alpha_R \sin(k) \sigma_y \tau_z$$
 altering Eq.~(\ref{eq:HQ1D}) accordingly.
In the two-chain problem, the two distinct sets of MKPs may interact, each hybridizing to finite energies. We note however, that the Hamiltonian (Eq. \ref{eq:HQ1D}) commutes with the spatial reflection operator ${\cal{R}}=\sigma_0 \tau_0 \rho_x$ which interchanges the chain index,
i.e. $\hat{c}_{k_x,1(2)}\mapsto \hat{c}_{k_x,2(1)}$.
Writing Eq.~(\ref{eq:HQ1D}) in the eigen-basis of ${\cal{R}}$ results in a block-diagonalized form which reads ${\cal{H}}^{1D}_k \rho_0+t_\perp \sigma_0 \tau_z \rho_z$. In this form it is clear that transverse hopping modifies the effective chemical potential in two independent bands.
Notice also that $\left[\Theta,{\cal R}\right]=\left[\Xi,{\cal R}\right]=0$, so that every diagonal block in the eigenbasis of ${\cal R}$ is particle-hole and time-reversal invariant.
Because of this invariance each independent, non-interacting block constitutes a DIII topological superconductor hosting a zero energy MKP at each end.
The extension of this argument to decouple $N_y$ chains is straightforward. A generalized $\rho_x$ is a totally symmetric $N_y \times N_y$ dimensional matrix given by $\rho_x=(\delta_{l,l'+1}+\delta_{l,l'-1})$ with $l,l' \in (1,2,...,N_y)$, that is, the superdiagonal and subdiagonal elements connect nearest neighbor sites are $+1$ and all other matrix elements are zero.
The eigenvalues of $\rho_x$ come in pairs of equal magnitude and opposite sign $(\pm \lambda_1,\pm \lambda_2,\pm \lambda_3,...)$ for $N_y$ even and $(0,\pm \lambda_1,\pm \lambda_2,\pm \lambda_3,...)$ when $N_y$ is odd.
In this case the rotated $N_y$ chain Hamiltonian involves a generalized $\rho_z$.
Because $\rho_z$ and $\rho_x$ have the same eigenvalue spectrum, the block digaonal Hamiltonian consists of non-interacting sectors where the chemical potential in sector is modified by $\pm \lambda_i$.

\textit{Persistence of Majorana multiplets in the absence of reflection symmetry:}
\begin{figure}[t!]
\includegraphics[width=8cm]{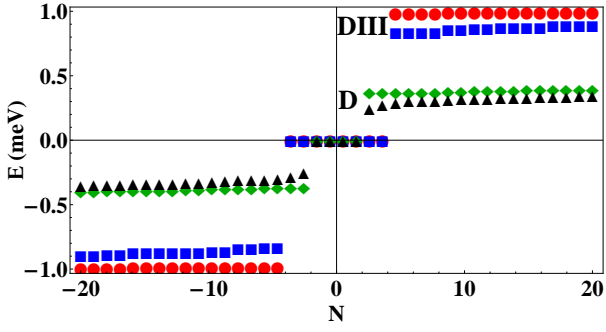}
\caption{\label{fig:Broken_Reflection_BdG}
(Color online) Low energy BdG quasiparticle spectrum for TR-symmetric Kitaev system (class DIII superconductor) for $N_y=2$ (red circles) in the absence of chiral and mirror symmetries. The eight MFs (four on each end) are protected from splitting by spatial reflection. Blue squares show same number of protected zero modes in the presence of local chemical potential disorder which breaks spatial reflection. Green diamonds and black triangles show two MFs at each end for
class D, $N_y=2$, systems with or without spatial reflection, respectively.}
\end{figure}
In the previous section we discussed the role of spatial reflection symmetry in protecting degenerate Majorana modes in a multi-chain setup.
Reflection symmetry however is only approximate since some disorder will always be present in any realistic system.
The addition of $\delta \mu_i$, with a random magnitude within a normal distribution, to the tight-binding Hamiltonian in Eq.\ref{eq:H1D} effectively models local disorder.
As illustrated in Fig. \ref{fig:Broken_Reflection_BdG}, by numerically solving the BdG equations on a double chain system, we find that the presence of on-site disorder minimally affects the bulk bandstructure, while the zero-energy modes are insensitive to this perturbation. Note that the Majorana multiplets persist even in the absence of chiral, mirror, and reflection symmetries, all of which are now explicitly broken. We get similar results even for $N_y>2$, and the number of Majorana multiplets scale with the number of chains in the transverse direction.

In order to understand the response of the MFs to reflection breaking perturbations we consider first
the two-chain Hamiltonian describing the chemical potential imbalance written as,
\begin{equation}
H^{Q1D}_{k}=
\left(
\begin{array}{cc}
 {\cal{H}}^{1D}_k+\delta  & -t_\perp\\
 -t_\perp & {\cal{H}}^{1D}_k-\delta
\end{array}
\right)
\label{eq:unequal_chains}
\end{equation}
where $t_\perp$ is understood to be $t_\perp \sigma_0 \tau_z$, $\delta=\delta \mu \sigma_0 \tau_z$, and we continue to work in the $(\Psi^{\dagger}_{k,1},\Psi^{\dagger}_{k,2})$ basis. Note that we still consider $k=k_x$ to be a good quantum number and break reflection symmetry only by introducing a chemical potential imbalance among the chains.
The question is now the following: Can we systematically block-diagonalize this Hamiltonian with a unitary eigenvalue-conserving transformation that commutes with time reversal and particle-hole symmetries? If yes, MFs will persist in each block due to a `hidden symmetry' associated with this transformation. Since ${{\cal{H}}_k}^{1D}$ appears with an identity matrix in chain space, this problem amounts to finding a matrix which diagonalizes the remaining terms leaving ${{\cal{H}}_k}^{1D}$ invariant.

We search for a hidden unitary transformation in a systematic way by first considering the eigen-decomposed form of the non-diagonal terms in Eq.~(\ref{eq:unequal_chains}), which we call $A=\delta \rho_z - t_\perp \rho_x$. This is expressed as $A = Q \Lambda  Q^{-1}$, where $Q$ is a matrix whose columns are the eigenvectors of $A$, $\bm{v}_\pm=1/(\sqrt{2}N_{\pm})\left(-(\delta \pm \sqrt{t^{2}_{\perp} +\delta^2})/t_{\perp},1 \right)^T$ and $N_{\pm}=\sqrt{1+\frac{\delta ^2}{t^2} \pm \frac{\delta  \sqrt{t^2+\delta ^2}}{t^2}}$ is the normalization constant. Also remember that each entry in $Q$ involves an identity in spin and particle-hole spaces. The unitarity of $Q$ is a direct consequence of the Hermiticity of $A$.
Note that in the limit $\delta \rightarrow 0$, this reduces to the eigenbasis of ${\cal{R}}$ which was the reflection transformation operator used in the presence of ${\cal R}$.
Rotating the full $8\times8$ Hamiltonian in Eq.~(\ref{eq:unequal_chains}) by the $Q$ operator we see $Q^{-1} H^{Q1D}_k Q = [-(\epsilon_k-\mu)\sigma_0 \tau_x+ \alpha^{R}_k \sigma_y \tau_z +\Delta \sigma_x \tau_x]\rho_0 -\sqrt{t_\perp^2+\delta \mu^2} \sigma_0 \tau_z \rho_z$.
The transformed Hamiltonian consists of two non-interacting topolgical DIII sectors, each block respecting both particle-hole and time reversal symmetries, subject to a modified chemical potential of magnitude $\sqrt{t^2_\perp +\delta^2}$ and a sign change for the single particle kinetic energy.
The commuting hidden symmetry operator assocaited with the $Q$ transformation is $R^{'}=\left(1+\frac{\delta \mu ^2}{t_{\perp}{}^2}\right){}^{-1/2}\left(\rho _x-\frac{\delta \mu }{t}\rho _z\right)$.
This explains why the multiple MKPs with spatially overlapping wave functions persist even with broken reflection symmetry, as shown in
Fig.~\ref{fig:Broken_Reflection_BdG} (but only as long as $k_x$ is a good quantum number).
What is the fate of the topological phase that hosts the Majorana modes in the presence of a perturbation ${H}_p$ that breaks the reflection symmetry and also the translation symmetry along the chains, a likely scenario in experimental systems due to disorder? Our strategy is to identify the generic structure of the Majorana wave functions $\phi_\nu$ based on the symmetries of the system, then calculate the matrix elements $\langle \phi_\nu|{H}_p|\phi_{\nu^\prime}\rangle$. If all matrix elements are zero, the Majorana multiplet is preserved; otherwise, the perturbation splits the Majorana modes and the systems becomes topologically trivial. Note that virtual transitions to finite energy states do not affect the energy of the zero-modes. This can be seen by writing the Green's function $G=(\omega - H^{Q1D} -H_p)^{-1}$ projected onto the Majorana subspace as $G_{\nu\nu^\prime} = [\omega\delta_{\nu\nu^\prime} - \Sigma_{\nu\nu^\prime}(\omega)]^{-1}$, in terms of the self-energy
\begin{equation}
\Sigma_{\nu\nu^\prime}(\omega) =\sum_n \frac{1}{E_n}\langle \phi_\nu|{H}_p|\Psi_n\rangle \langle \Psi_n|{H}_p|\phi_{\nu^\prime}\rangle,
\end{equation}
where $\Psi_n$ is an eigenstate of energy $E_n$. Since $|E_n|\geq \Delta_{qp}$, where $\Delta_{qp}$ is the quasiparticle gap,  $\Sigma_{\nu\nu^\prime}=0$ in the limit $\Delta_{qp}\rightarrow\infty$. Furthermore, any system with a finite quasiparticle gap  can be adiabatically connected to the superconductor with infinite $\Delta_{qp}$ without crossing a topological phase transition; hence, the two systems belong to the same topological class and have the same number of Majorana modes, i.e.  $\Sigma_{\nu\nu^\prime}=0$ for both. Explicit numerical calculations confirm this result.

To identify the generic form of the Majorana wave function, we make the key observation that  $H^{Q1D}$ commutes with the unitary operator $U=\sigma_y\tau_z\rho_0$. The Majorana spinor $\phi_\nu = (u_{\nu\uparrow}, u_{\nu\downarrow}, v_{\nu\downarrow}, -v_{\nu\uparrow})^T$, which satisfies the constraint $v_{\nu\sigma} = u_{\nu\sigma}^*e^{i \varphi}$ due to particle-hole symmetry, has to be an eigenstate of $U$. Consequently,
$u_{\nu\downarrow} =i \lambda u_{m\lambda\uparrow}$, where $\nu=(m, \lambda)$, $\lambda=\pm 1$ and $m$ takes $N_y$ values to label the Majorana multiplet localized near each end of the system. Finally, time-reversal symmetry requires $u_{m\lambda\uparrow}=u_{m\lambda}(l, i)$ to be a real function of position and chain index satisfying the condition $u_{m +}=u_{m -}$, in addition to the phase condition $\varphi|_{\lambda=+1} = -\varphi|_{\lambda=-1}$.  We conclude that the Majorana wave functions have the generic form
\begin{equation}
\phi_{m\lambda}(l, i) = u_{m\lambda}(l, i) [1, \lambda i, -\lambda i e^{\lambda i \varphi}, -e^{\lambda i \varphi}]^T.   \label{Mfunc}
\end{equation}
We note that the phase $\varphi$ takes the value $\varphi=\pi$ in a system with chiral symmetry, i.e. in the absence of spin-orbit coupling ($\alpha^R=0$). Using Eq. (\ref{Mfunc}), it is straightforward to show that all matrix elements $\langle \phi_{m\lambda}|{H}_p|\phi_{m^\prime\lambda^\prime}\rangle$ of a spin-independent disorder potential $H_p = V_{dis}(l,i) \sigma_0\tau_z$ vanish; hence, such a perturbation does not destroy the Majorana multiplet as long as the quasiparticle gap is nonzero. This explains the numerical BdG results and persistence of the end MFs presented in Fig.~\ref{fig:Broken_Reflection_BdG}.

\textit{Conclusion:}
Multiple Majorana fermions with spatially overlapping wave functions are expected to split and acquire non-zero energies, unless such splitting breaks an underlying symmetry. By working with a realistic model for TR-invariant topological superconductors (class DIII) appropriate for LiMO, we study such interaction effects in MF multiplets and show that they can remain protected in topological superconductors even in the absence of an identifiable physical symmetry. We find similar results for Q1D semiconductor-superconductor heterostructures with spin-orbit coupling and Zeeman field (class D). Our results, which we explain in terms of properties of the Hamiltonians and wave functions, underscore the importance of hidden symmetry decoupling of MFs in topological superconductors.

{\em Acknowledgment:} Work supported by  NSF (PHY-1104527), AFOSR (FA9550-13-1-0045), and WV HEPC/dsr.12.29.


\end{document}